\documentclass[prd,showpacs,preprintnumbers]{revtex4} % Physical Review D

\usepackage{graphicx}% Include figure files
\usepackage{dcolumn}% Align table columns on decimal point
\usepackage{bm}% bold math
\usepackage{latexsym}
\usepackage{amsfonts}
\usepackage{amssymb}
\usepackage{amsmath}
\usepackage[usenames]{color}
\begin{document}

\title{A Note on Gauss-Bonnet Holographic Superconductors}

\author{Sugumi Kanno}
\affiliation{Institute of Cosmology, Department of Physics and Astronomy, 
Tufts University, Medford, Massachusetts 02155, USA
}

\date{\today}% It is always \today, today,
             %  but any date may be explicitly specified

%===============================================================%
%************************* ABSTRACT ****************************%
%===============================================================%
\begin{abstract}
We present an analytic treatment near the phase
transition for the critical temperature of (3+1)-dimensional 
holographic superconductors in Einstein-Gauss-Bonnet gravity 
with backreaction. We find that the backreaction makes the 
critical temperature of the superconductor decrease and 
condensation harder.
This is consistent with previous numerical results.
\end{abstract}

\pacs{11.25.Tq,04.70.Bw,74.20.-z}
% PACS, the Physics and Astronomy
% Classification Scheme.
%\keywords{Suggested keywords}
%Use showkeys class option if keyword
%display desired
\maketitle

\section{Introduction}

Gauge/Gravity duality stems from string theory
provides a rich tool for analyzing strongly coupled
 field theory~\cite{Maldacena:1997re}. 
Especially, the duality provides an established method for 
calculating correlation functions in a strongly 
interacting field theory using 
a dual classical gravity description~\cite{Gubser:1998bc, Witten:1998qj}. 
Holographic superconductors established in~\cite{HHH1,HHH2}
are remarkable examples where the Gauge/Gravity
duality plays an important role.
There, a superconducting phase transition is
described by black hole physics according to the duality.

Many of previous works on the holographic superconductors were
performed in the probe limit, where the backreaction of matter fields 
on the spacetime metric is neglected.
However, we found that the effect of the Gauss-Bonnet coupling lowers
the critical temperature of holographic superconductors in previous
work~\cite{Gregory:2009fj}. Backreaction becomes important 
when considering lower temperature of 
black holes in AdS spacetime, because lower Hawking temperature of black
holes means smaller black holes, i.e., larger Coulomb energy 
of the matter fields near the black hole horizon.

The critical temperature was 
obtained numerically both with and without the 
backreaction~\cite{HHH1,HHH2,Horowitz:2008bn}. 
As an analytic approach for deriving the critical temperature, 
an approximate analytic formula was proposed in the probe limit 
by a matching method~\cite{Gregory:2009fj}.
An alternative analytic method using the expansion
around the critical point 
where the phase transition occurs
was also proposed~\cite{Herzog:2010vz}. Some other analytic approaches
were also proposed in the probe limit~\cite{Siopsis:2010uq, Li:2011xj, 
Ge:2010aa, Pan:2009xa}.
In this note, we derive analytically the critical temperature
of the Gauss-Bonnet holographic superconductors with backreaction
by combining the small backreaction approximation and the matching method
near the critical point.

\section{Gauss-Bonnet Black Holes}
We begin with the action for a Maxwell field and a charged complex scalar field
coupled to the Einstein-Gauss-Bonnet:
\begin{eqnarray}
S&=&\frac{1}{2\kappa^2}\int d^5x \sqrt{-g}
\left[
R+\frac{12}{L^2}+\frac{\alpha}{2}\left(
R^{\mu\nu\lambda\rho}R_{\mu\nu\lambda\rho}
-4R^{\mu\nu}R_{\mu\nu}+R^2
\right)
\right]\nonumber\\
&&\hspace{3.5mm}+\int d^5x\sqrt{-g}\left[
-\frac{1}{4}F^{\mu\nu}F_{\mu\nu}-|\nabla\psi -iqA\psi|^2
-m^2|\psi|^2
\right]\,,
\label{action}
\end{eqnarray}
where $g$ is the determinant of a metric $g_{\mu\nu}$ and
$R_{\mu\nu\lambda\rho}$, $R_{\mu\nu}$ and $R$ are the Riemann curvature tensor,
Ricci tensor, and the Ricci scalar, respectively. $q$ is the charge of the
scalar field.
We take the Gauss-Bonnet coupling constant $\alpha$ to be positive.
Here, the negative cosmological constant term $-6/L^2$ is also introduced. 
We look for electrically charged plane-symmetric hairy black hole solutions
taking the metric ansatz
\begin{eqnarray}
ds^2 = -f(r)e^{2\nu(r)}dt^2 + \frac{dr^2}{f(r)} 
+ \frac{r^2}{L^2}(dx^2+dy^2+dz^2)
\end{eqnarray}
together with a static ansatz for the fields,
\begin{eqnarray}
A_\mu=(\phi(r),0,0,0,0)\,,\hspace{1cm}\psi =\psi(r).
\end{eqnarray} 
Here, without loss of generality $\psi$ can be taken to be
real. First, we look for the solutions in normal phase, $\psi=0$.
We find $\nu$ is constant and 
\begin{eqnarray}
\phi=\mu - \frac{\rho}{r^2}\,.
\end{eqnarray}
Here, $\mu$ and $\rho$ are interpreted as a chemical
potential and charge density of the dual theory on the
boundary, respectively. 
We also find 
\begin{eqnarray}
f(r) = \frac{r^2}{2\alpha}\left[
1-\sqrt{1-\frac{4\alpha}{L^2}\left(
1-\frac{r_+^4}{r^4}
\right)
+2\kappa^2\frac{4\alpha\rho^2}{3r^4r_+^2}
\left(1-\frac{r_+^2}{r^2}
\right)}
\right]\,,
\label{sol}
\end{eqnarray} 
where we chose the minus sign of the solutions so that we have
a solution in the Einstein limit ($\alpha\rightarrow 0$). The 
horizon radius $r_+$ is defined through the requirement that
$f(r_+)=0$, so we set a constant of integration by imposing this 
condition at the horizon. We see the above solution becomes
the Reissner-Nordstr\"{o}m-AdS solution in the Einstein limit.
In order to avoid a naked
singularity, we need to restrict the parameter range as $\alpha\leq L^2/4$.
For general $\alpha$, the solution (\ref{sol}) behaves as
\begin{eqnarray}
f(r)\sim\frac{r^2}{2\alpha}\left[
1-\sqrt{1-\frac{4\alpha}{L^2}}
~\right]\,,
\end{eqnarray}
in the asymptotic region.
Hence, we define the effective asymptotic AdS scale by
\begin{eqnarray}\label{Leff}
L^2_{\rm e}=\frac{2\alpha}{1-\sqrt{1-\frac{4\alpha}{L^2}}}
\to  \left\{
\begin{array}{rl}
L^2   \ , &  \quad {\rm for} \ \alpha \rightarrow 0 \\
\frac{L^2}{2}  \ , &  \quad {\rm for} \  \alpha \rightarrow \frac{L^2}{4}
\end{array}\right.
\,.
\end{eqnarray}
Here, the maximum value of $\alpha$ is called the Chern-Simons limit.
The Hawking temperature is given by
\begin{eqnarray}
T_H = \frac{1}{4\pi} f' (r)e^{\nu(r)}\bigg|_{r=r_+}\ ,
\label{hawking}
\end{eqnarray}
where a prime denotes derivative with respect to $r$.
This will be interpreted as the temperature of the holographic superconductors.

\section{Gauss-Bonnet Holographic Superconductors}

In order to obtain the solutions in superconducting phase, $\psi\neq 0$, 
we need to take into account the boundary conditions
at the horizon and the AdS boundary.
The position of the horizon, $r_+$, is defined through $f(r_+)=0$. 
We can set $\nu(r_+)=0$ there by rescaling the time coordinate.
Then the Einstein equations give:
\begin{eqnarray}
&&
\nu^\prime(r_+)=\frac{2\kappa^2}{3}r_+\left(
\psi^\prime(r_+)^2
+\frac{q^2\phi^\prime(r_+)^2\psi(r_+)^2}{f^\prime(r_+)^2e^{2\nu(r_+)}}
\right)\,,\\
&&
f^\prime(r_+)=\frac{4}{L^2}r_+
-\frac{2\kappa^2}{3}r_+\left(
\frac{\phi^\prime(r_+)^2}{2e^{2\nu(r_+)}}
+m^2\psi(r_+)^2
\right)\,.
\end{eqnarray}

\noindent $\bullet$ Regularity at the horizon for $\phi$ and $\psi$ 
gives two conditions:
\begin{eqnarray}
\phi(r_+)=0,\hspace{1cm}
\psi^\prime(r_+)=\frac{m^2}{f^\prime(r_+)}\psi(r_+) \ .
\end{eqnarray}
As we want the spacetime to be asymptotically AdS, we look for a solution
with 
\begin{eqnarray}
&&\nu(r)={\rm const.}\,,\hspace{1cm}
f(r)= \frac{r^2}{L_e^2}\,,
\end{eqnarray}
at the AdS boundary.

\noindent $\bullet$ Asymptotically ($r\rightarrow\infty$) the 
solutions of $\phi$ and $\psi$ are found to be:
\begin{eqnarray}
\phi(r)\sim\mu - \frac{\rho}{r^2}\,,\hspace{1cm} 
\psi\sim\frac{C_{-}}{r^{\Delta_-}}+\frac{C_{+}}{r^{\Delta_+}}\,,
\label{r:boundary}
\end{eqnarray}
where $\Delta_\pm=2\pm\sqrt{4+m^2L_e^2}$. 
Note that these are
not entirely free parameters, as there is a scaling degree of 
freedom in the equations of motion. As in \cite{HHH1}, we impose 
that $\rho$ is fixed, which determines the scale of the system. 
For $\psi$, in order to have a normalizable solution
we take $C_{-}=0$.

According to the Gauge/Gravity duality
~\cite{Maldacena:1997re,Gubser:1998bc,Witten:1998qj}, 
we can interpret 
$ \langle {\cal O}_{\Delta_+} \rangle \equiv C_{+}$, 
where ${\cal O}_{\Delta_+}$ is the operator with the conformal
dimension $\Delta_+$ dual to the scalar field. 
Thus, we are going to calculate the condensate 
$\langle {\cal O}_{\Delta_+} \rangle$ 
for fixed charge density.
We note that 
since $C_{-}$ is regarded as the source term of the
operator, we put it zero at the end by using the Gauge/Gravity 
dictionary~\cite{Gubser:1998bc,Witten:1998qj},
which is consistent with taking $C_{-}=0$.

Let us change the coordinate and set $z=r_+/r$. Under this
transformation, the Einstein, Maxwell and the scalar equations become
\begin{eqnarray}
&&\left(1-2\alpha\frac{z^2}{r_+^2}f\right)\nu^\prime=
-\frac{2\kappa^2}{3}\frac{r_+^2}{z^3}\left(
\frac{q^2\phi^2\psi^2}{f^2e^{2\nu}}+\frac{z^4}{r_+^2}\psi^{\prime2}
\right)\,,
\label{z:nu}\\
&&\left(1-2\alpha\frac{z^2}{r_+^2}f\right)f^\prime
-\frac{2}{z}f+\frac{4r_+^2}{L^2z^3}=\frac{2\kappa^2}{3}\frac{r_+^2}{z^3}
\left[\frac{z^4}{2r_+^2e^{2\nu}}\phi^{\prime2}
+m^2\psi^2+f\left(\frac{q^2\phi^2\psi^2}{f^2e^{2\nu}}
+\frac{z^4}{r_+^2}\psi^{\prime2}
\right)\right]\,,
\label{z:f}\\
&&\phi^{\prime\prime}-\left(\frac{1}{z}+\nu^\prime\right)\phi^\prime
-\frac{2r_+^2}{z^4}\frac{\psi^2}{f}\phi=0\,,
\label{z:phi}\\
&&\psi^{\prime\prime}-\left(
\frac{1}{z}-\nu^\prime-\frac{f^\prime}{f}\right)\psi^\prime
+\frac{r_+^2}{z^4}\left(
\frac{\phi^2}{f^2e^{2\nu}}-\frac{m^2}{f}
\right)\psi=0\,,
\label{z:psi}
\end{eqnarray}
where the prime now denotes a derivative with respect to $z$. 
The region $r_+<r<\infty$ now corresponds to $0<z<1$.
If one sets $\tilde{\phi}=\phi/q$, $\tilde{\psi}=\psi/q$ in
the action~(\ref{action}),
the Maxwell and the scalar equations remain unchanged, while
the gravitational coupling of the Einstein equations changes
$\kappa^2\rightarrow\kappa^2/q^2$. If one takes the limit
$q\rightarrow\infty$, the matter sources drop out of the
Einstein equations and this is the probe limit. To go beyond
the probe limit, we can take either finite $q$ with setting
$2\kappa^2=1$, or finite $2\kappa^2$ with setting $q=1$. The
paper~\cite{HHH2} took the former choice to consider the 
effects of backreaction of the spacetime metric,
but we will take the latter choice in the following.

In order to solve these equations, we focus on near the critical point
as in~\cite{Maeda:2008ir, Herzog:2010vz}, around which the
stability was confirmed~\cite{Kanno:2010pq}.
It is convenient to introduce a scalar operator as an expansion 
parameter:
\begin{eqnarray}
\epsilon\equiv\langle{\cal O}_{\Delta_+}\rangle\,.
\end{eqnarray}
As $\psi$ is small near the critical point, we expand $\psi$ from
the first order. From Eq.~(\ref{z:phi}), $\phi$ and $\psi$ are expanded by
$\epsilon^2$ subsequently as follows:
\begin{eqnarray}
\phi&=&\phi_0+\epsilon^2\phi_2+\epsilon^4\phi_4+\cdots\,,\\
\psi&=&\epsilon\psi_1+\epsilon^3\psi_3+\epsilon^5\psi_5+\cdots\,,
\end{eqnarray}
and in this situation where starting from the normal phase, 
the background can be expanded around the 
Reissner-Nordst{\"o}m-AdS spacetime:
\begin{eqnarray}
f&=&f_0+\epsilon^2f_2+\epsilon^4f_4+\cdots\,,\\
\nu&=&\epsilon^2\nu_2+\epsilon^4\nu_4\cdots\,,
\end{eqnarray}
where $\epsilon\ll 1$. 

While $\mu$ is also expanded by the order parameter as
\begin{eqnarray}
\mu=\mu_0 + \epsilon^2\delta\mu_2\,,
\end{eqnarray}
where $\delta\mu_2$ is also positive.
So, we find the order parameter as a function of
the chemical potential, which is expressed by
\begin{eqnarray}
\epsilon=\langle{\cal O}_{\Delta_+}\rangle
=\left(\frac{\mu-\mu_0}{\delta\mu_2}\right)^{1/2}\,.
\end{eqnarray}
The exponent $1/2$ is consistent with the Ginzburg-Landau mean field theory 
for phase transitions. When $\mu=\mu_0$,
the order parameter becomes zero, which means the critical value of $\mu$ is
defined by $\mu_0$ such as $\mu_c=\mu_0$.
Now we begin to solve equations order by order.

At zeroth order, we impose $\phi_0(1)=0$ at the horizon, $z=1$, and 
$\phi_0(0)=\mu_0={\rm const.}$ at the boundary, $z=0$.
Then the solution of Eq.~(\ref{z:phi}) is given by
\begin{eqnarray}
\phi_0(z)=\mu_0(1-z^2)\,.
\label{0:phi}
\end{eqnarray}
This gives a relation $\rho=\mu_0 r_+^2$ by the coordinate transformation. 
Inserting this solution into Eq.~(\ref{z:f}), we obtain
\begin{eqnarray}
f_0(z)=\frac{r_+^2}{2\alpha}\frac{1}{z^2}\left[
1-\sqrt{1-\frac{4\alpha}{L^2}(1-z^4)
+2\kappa^2\frac{4\alpha\mu_0^2}{3r_+^2}z^4(1-z^2)}
~\right]\,,
\end{eqnarray}
where we chose the minus sign of the solutions so that we have a
solution in the Einstein limit. We also used $f_0(1)=0$ at the horizon
as we did in Eq.~(\ref{sol}). 
We find the above solution becomes
the Reissner-Nordstr\"{o}m-AdS solution in the Einstein limit.

Even at first order, it is difficult to deal with both of 
the backreaction on the spacetime metric and the Gauss-Bonnet term. 
Here we will appeal to the matching method used in~\cite{Gregory:2009fj}.
Before proceeding to it, let us check the regularity condition at 
the horizon at this order:
\begin{eqnarray}
\psi_1^\prime(1)=\frac{r_+^2m^2}{f_0^\prime(1)}\psi_1(1)
\label{1:regularity} \ .
\end{eqnarray}
The behavior of $\psi$ at the boundary is given by
\begin{eqnarray}
\psi_1\sim D_-z^{\Delta_-}+D_+z^{\Delta_+}\,,
\label{1:boundary}
\end{eqnarray}
where $\Delta_\pm$ is the same as in Eq.~(\ref{r:boundary}). 
For the boundary conditions, we take $D_-$ to be zero as we did 
in Eq.~(\ref{r:boundary}). 

Now we find the solution of $\psi_1$ under these conditions by 
the matching method and derive the critical temperature.
We can expand $\psi_1$ in a Taylor series near the horizon as:
\begin{eqnarray}
\psi_1(z)=\psi_1(1)-\psi_1^\prime(1)(1-z)
+\frac{1}{2}\psi_1^{\prime\prime}(1)(1-z)^2
+\cdots
\end{eqnarray}
Here, we have Eq.~(\ref{1:regularity}) for the first order coefficients of 
$\psi_1$, and without loss of generality
we can take $\psi_1(1)>0$ to have $\psi_1(z)$ positive. 
The second order coefficients of $\psi_1$ are computed by using 
Eq.~(\ref{z:psi}). Then we find the second derivative at the horizon
is expressed by
\begin{eqnarray}
\psi_1^{\prime\prime}(1)=\frac{1}{2}\left(
-3-\frac{f_0^{\prime\prime}(1)}{f_0^\prime(1)}+\frac{r_+^2m^2}{f_0^\prime(1)}
\right)\psi_1^\prime(1)
-\frac{r_+^2\phi_0^\prime(1)^2}{2f_0^\prime(1)^2}\psi_1(1)\,.
\end{eqnarray}
After eliminating $\psi_1^\prime(1)$ from above equation 
by using Eq.~(\ref{1:regularity}), an approximate solution near the 
horizon is given by
\begin{eqnarray}
\psi_1(z)=\psi_1(1)-\frac{r_+^2m^2}{f_0^\prime(1)}\psi_1(1)(1-z)
+\left[-\frac{r_+^2m^2}{4f_0^\prime(1)}\left(
3+\frac{f_0^{\prime\prime}(1)}{f_0^\prime(1)}
-\frac{r_+^2m^2}{f_0^\prime(1)}
\right)-\frac{r_+^2}{4}\frac{\phi_0^\prime(1)^2}{f_0^\prime(1)^2}
\right]\psi_1(1)(1-z)^2
+\cdots
\label{1:horizon}
\end{eqnarray} 
Here, $\psi_1(1)$ is still unknown.

On the other hand, from (\ref{1:boundary}), $\psi_1$ in the asymptotic 
region are given by
\begin{eqnarray}
\psi_1(z)\sim D_+z^{\Delta_+}\,.
\label{1:boundary2}
\end{eqnarray}
We have set $D_-=0$ from the boundary condition. Here $D_+$ is another
unknown constant.

Now we try to connect the solutions Eq.~(\ref{1:horizon}) and 
(\ref{1:boundary2}) smoothly at $z_m$ in order to obtain 
$\psi_1(1)$ and $D_+$.
As we shall see below the choice of $z_m$
does not change the qualitative features of solutions.
In order to 
connect those solutions smoothly, we require the following two 
conditions: 
\begin{eqnarray}
&&z_m^{\Delta_+}D_+
=\psi_1(1)-\frac{r_+^2m^2}{f_0^\prime(1)}(1-z_m)\psi_1(1)
+\left[
-\frac{r_+^2m^2}{4f_0^\prime(1)}\left(
3+\frac{f_0^{\prime\prime}(1)}{f_0^\prime(1)}
-\frac{r_+^2m^2}{f_0^\prime(1)}
\right)
-\frac{r_+^2}{4}\frac{\phi_0^\prime(1)^2}{f_0^\prime(1)^2}
\right](1-z_m)^2\psi_1(1)
\label{m:psi}\,,\\
&&\Delta_+ z_m^{\Delta_+-1} D_+
=\frac{r_+^2m^2}{f_0^\prime(1)}\psi_1(1)
-2\left[
-\frac{r_+^2m^2}{4f_0^\prime(1)}\left(
3+\frac{f_0^{\prime\prime}(1)}{f_0^\prime(1)}
-\frac{r_+^2m^2}{f_0^\prime(1)}
\right)
-\frac{r_+^2}{4}\frac{\phi_0^\prime(1)^2}{f_0^\prime(1)^2}
\right](1-z_m)\psi_1(1)\,.
\label{m:dpsi}
\end{eqnarray}
From Eqs~(\ref{m:psi}) and (\ref{m:dpsi}), we find the relation
between $\psi_1(1)$ and $D_+$,
\begin{eqnarray}
D_+=\frac{2z_m}{2z_m+(1-z_m)\Delta_+}z_m^{-\Delta_+}\left(
1-\frac{1-z_m}{2}\frac{r_+^2m^2}{f_0^\prime(1)}
\right)\psi_1(1)\,.
\label{D}
\end{eqnarray}
Plugging this back into Eq.~(\ref{m:dpsi}), we find the following
relation in order to get a non-trivial solution, $\psi_1(1)\neq 0$, 
\begin{eqnarray}
&&\frac{2\Delta_+}{2z_m+(1-z_m)\Delta_+}
-\left(\frac{(1-z_m)\Delta_+}{2z_m+(1-z_m)\Delta_+}
+\frac{5-3z_m}{2}
\right)\frac{r_+^2m^2}{f_0^\prime(1)}
\nonumber\\
&&\hspace{3.1cm}
-\frac{(1-z_m)r_+^2m^2}{2}\frac{f_0^{\prime\prime}(1)}{f_0^\prime(1)^2}
+\frac{1-z_m}{2}\frac{r_+^4m^4}{f_0^\prime(1)^2}
-\frac{(1-z_m)r_+^2}{2}\frac{\phi_0^\prime(1)^2}{f_0^\prime(1)^2}
=0
\label{nontrivial}\,.
\end{eqnarray}

Similarly, plugging Eq.~(\ref{D}) back in Eq.~(\ref{m:psi}) we can
get the solution of $\psi_1(1)$ and then $D_+$ as well in principle.
But what we want to know here is the critical temperature rather than
deriving the solution of $\psi$ up to this order, so we are
going to focus on it in the following. For this purpose, putting the values 
of $f_0^\prime(1)$, $f_0^{\prime\prime}(1)$ and
$\phi_0^\prime(1)$ in the above relation, then Eq.~(\ref{nontrivial})
yields the equation for $\mu_0$:
\begin{eqnarray}
&&4\kappa^4\frac{L^4}{36r_+^4}\left[
\frac{2\Delta_+}{2z_m+(1-z_m)\Delta_+}-(1-z_m)m^2\alpha
\right]\mu_0^4
\nonumber\\
&&
-\frac{(1-z_m)L^4}{8r_+^2}\left[
1+2\kappa^2\left\{\left(
\frac{16}{3(1-z_m)L^2}+\frac{m^2}{3}
\right)\frac{\Delta_+}{2z_m+(1-z_m)\Delta_+}
+\left(\frac{5-3z_m}{6(1-z_m)}+\frac{5}{6}\right)m^2
-\frac{8m^2}{3L^2}\alpha\right\}
\right]\mu_0^2\nonumber\\
&&+\frac{(1-z_m)\Delta_+}{2z_m+(1-z_m)\Delta_+}\left(
\frac{2}{1-z_m}+\frac{m^2L^2}{4}\right)
+\frac{2-z_m}{4}m^2L^2+\frac{1-z_m}{32}m^4L^4
-(1-z_m)m^2\alpha=0\,.
\end{eqnarray}

In order to solve the above equation with respect to $\mu_0$, 
we assume $\kappa^2\ll 1$ in the following. 
This means that all functions are expanded
by $\kappa^2$ as well. That is, solutions near the phase transition
are obtained in the
small backreaction approximation together with the matching method. 
As the first term disappears from the above equation,
 we can solve it easily and we get
\begin{eqnarray}
\mu_0&=&\sqrt{\frac{8}{1-z_m}}\frac{r_+}{L^2}\left[
\frac{(1-z_m)\Delta_+}{2z_m+(1-z_m)\Delta_+}\left(
\frac{2}{1-z_m}+\frac{m^2L^2}{4}\right)
+\frac{2-z_m}{4}m^2L^2+\frac{1-z_m}{32}m^4L^4
-(1-z_m)m^2\alpha
\right]^{1/2}
\nonumber\\
&&\times
\left[
1-2\kappa^2\left\{\left(
\frac{16}{3(1-z_m)L^2}+\frac{m^2}{3}
\right)\frac{\Delta_+}{4z_m+2(1-z_m)\Delta_+}
+\left(\frac{5-3z_m}{6(1-z_m)}+\frac{5}{6}\right)\frac{m^2}{2}
-\frac{4m^2}{3L^2}\alpha\right\}
\right]\,,
\label{mu0}
\end{eqnarray}
where $\mu_0$ is positive. Combining Eq.~(\ref{mu0}) with 
the relation $\mu_0=\rho/r_+^2$, which is given under Eq.~(\ref{0:phi}), 
we find $r_+$ is given by
\begin{eqnarray}
r_+&=&\frac{\rho^{1/3}}{\pi L^{4/3}}\left(
\frac{1-z_m}{8}
\right)^{1/6}\left[
\frac{~8\Delta_+ +(1-z_m)\Delta_+m^2L^2}
{8z_m+4(1-z_m)\Delta_+}
+\frac{2-z_m}{4}m^2L^2+\frac{1-z_m}{32}m^4L^4
-(1-z_m)m^2\alpha~
\right]^{-1/6}
\nonumber\\
&&\hspace{0.5cm}
\times\left[
1+\frac{2\kappa^2}{L^2}\left\{
\frac{\Delta_+}{12z_m+6(1-z_m)\Delta_+}\left(
\frac{m^2L^2}{3}+\frac{16}{3(1-z_m)}
\right)
+\frac{5-4z_m}{1-z_m}\frac{m^2L^2}{18}
-\frac{4m^2}{9}\alpha
\right\}
\right]\,.
\label{r+}
\end{eqnarray}

The Hawking temperature Eq.~(\ref{hawking}) up to this order 
becomes 
\begin{eqnarray}
T_H=-\frac{f_0^\prime(1)e^{\nu_0(1)}}{4\pi r_+}
=\frac{r_+}{\pi L^2}\left(
1-2\kappa^2\frac{L^2}{6r_+^2}\mu_0^2
\right)\,.
\end{eqnarray}
The critical temperature is defined at the point where the
order parameter becomes zero, which leads to $T_H=T_c$ at $\mu_0=\mu_c$.
Eliminating $\mu_0(=\mu_c)$ from the above temperature $T_H$
using Eq.~(\ref{mu0}), we get the critical temperature
\begin{eqnarray}
T_c=\frac{r_+}{\pi L^2}\left[
1-\frac{2\kappa^2}{L^2}\left\{
\frac{4\Delta_+}{6z_m+3(1-z_m)\Delta_+}\left(
\frac{2}{1-z_m}+\frac{m^2L^2}{4}\right)
+\frac{2-z_m}{3(1-z_m)}m^2L^2+\frac{m^4L^4}{24}
-\frac{4m^2\alpha}{3}
\right\}
\right]\,.
\end{eqnarray}
Substituting $r_+$ with Eq.~(\ref{r+}), we obtain the critical temperature
of the form
\begin{eqnarray}
T_c&=&T_1
%\nonumber\\
%&&\hspace{1cm}
%\times
\left(
1-\frac{2\kappa^2}{L^2}T_2
\right)\,,
\label{tc1}
\end{eqnarray}
where
\begin{eqnarray}
T_1&=&\frac{\rho^{1/3}}{\pi L^{4/3}}\left(
\frac{1-z_m}{8}
\right)^{1/6}\left[
\frac{~8\Delta_+ +(1-z_m)\Delta_+m^2L^2}
{8z_m+4(1-z_m)\Delta_+}
+\frac{2-z_m}{4}m^2L^2+\frac{1-z_m}{32}m^4L^4
-(1-z_m)m^2\alpha~
\right]^{-1/6}
\label{t1}\,,\\
T_2&=&\frac{2\Delta_+}{2z_m+(1-z_m)\Delta_+}\left(
\frac{5m^2L^2}{36}+\frac{8}{9(1-z_m)}
\right)+\frac{7-2z_m}{1-z_m}\frac{m^2L^2}{18}
+\frac{m^4L^4}{24}-\frac{8m^2\alpha}{9}
\label{t2}\,.
\end{eqnarray}

\section{Conclusion}

We presented an analytic treatment near the phase
transition for the critical temperature of $(3+1)$-dimensional
holographic superconductors in Einstein-Gauss-Bonnet gravity
with backreaction by matching the solution expanded from infinity
with that expanded from the horizon. In this method, there is an
ambiguity in the choice of the matching radius. However, the result
turns out to be fairly insensitive to the choice of it.
The result reproduces our previous analytic calculation for
$m^2=-3/L^2$, $\kappa^2=0$ with the choice $z_m=1/2$, which was
in good agreement with the numerical result~\cite{Gregory:2009fj}.
It also agrees with \cite{Herzog:2010vz} for $m^2=-4/L^2$, $\kappa^2=\alpha=0$
with the choice $z_m=1/2$. It may be noted that the choice $z_m\sim0.5$
is roughly equal to  $z=\sqrt{r_+/L}$, corresponding to
the geometrical mean of the horizon radius and the AdS scale,
 $r=\sqrt{r_{+}L}$.
We found that the coefficient of the corrections due to backreaction,
$T_2$, is positive definite irrespective of the value of $\alpha$.
This means the effects of the backreaction makes condensation
harder. This result also agrees with the numerical results
obtained in~\cite{Brihaye:2010mr,Barclay:2010up, Pan:2011ns}. 

%\begin{acknowledgements}
\acknowledgements
{I would like to thank Jiro Soda for useful and stimulating
discussions and Misao Sasaki for helpful comments and suggestions.
I would also like to thank Luke Barclay, Ruth Gregory and Paul Sutcliffe
for previous collaboration and useful discussions. I was a member of 
the CPT group at the Department of Mathematical Sciences,
Durham University, and was supported by an STFC rolling grant while
part of this work was carried out. This work is supported in part 
by grant PHY-0855447 from the National Science 
Foundation.
}%\end{acknowledgements}


\begin{thebibliography}{99}

%\cite{Maldacena:1997re}
\bibitem{Maldacena:1997re}
  J.~M.~Maldacena,
  %``The Large N limit of superconformal field theories and supergravity,''
  Adv.\ Theor.\ Math.\ Phys.\  {\bf 2}, 231-252 (1998).
  [hep-th/9711200].

%\cite{Gubser:1998bc}
\bibitem{Gubser:1998bc}
  S.~S.~Gubser, I.~R.~Klebanov, A.~M.~Polyakov,
  %``Gauge theory correlators from noncritical string theory,''
  Phys.\ Lett.\  {\bf B428}, 105-114 (1998).
  [hep-th/9802109].

%\cite{Witten:1998qj}
\bibitem{Witten:1998qj}
  E.~Witten,
  %``Anti-de Sitter space and holography,''
  Adv.\ Theor.\ Math.\ Phys.\  {\bf 2}, 253-291 (1998).
  [hep-th/9802150].

\bibitem{HHH1}
%\cite{Hartnoll:2008vx}
%\bibitem{Hartnoll:2008vx}
  S.~A.~Hartnoll, C.~P.~Herzog and G.~T.~Horowitz,
  %``Building a Holographic Superconductor,''
  Phys.\ Rev.\ Lett.\  {\bf 101}, 031601 (2008)
  [arXiv:0803.3295 [hep-th]].
  %%CITATION = PRLTA,101,031601;%%

\bibitem{HHH2}
%\cite{Hartnoll:2008}
%\bibitem{Hartnoll:2008kx}
  S.~A.~Hartnoll, C.~P.~Herzog and G.~T.~Horowitz,
  %``Holographic Superconductors,''
  JHEP {\bf 0812}, 015 (2008)
  [arXiv:0810.1563 [hep-th]].
  %%CITATION = JHEPA,0812,015;%%

%\cite{Gregory:2009fj}
\bibitem{Gregory:2009fj}
  R.~Gregory, S.~Kanno and J.~Soda,
  %``Holographic Superconductors with Higher Curvature Corrections,''
  JHEP {\bf 0910}, 010 (2009)
  [arXiv:0907.3203 [hep-th]].
  %%CITATION = JHEPA,0910,010;%%

%\cite{Horowitz:2008bn}
\bibitem{Horowitz:2008bn}
  G.~T.~Horowitz and M.~M.~Roberts,
  %``Holographic Superconductors with Various Condensates,''
  Phys.\ Rev.\  D {\bf 78}, 126008 (2008)
  [arXiv:0810.1077 [hep-th]].
  %%CITATION = PHRVA,D78,126008;%%

%\cite{Herzog:2010vz}
\bibitem{Herzog:2010vz}
  C.~P.~Herzog,
  %``An Analytic Holographic Superconductor,''
  Phys.\ Rev.\  D {\bf 81}, 126009 (2010)
  [arXiv:1003.3278 [hep-th]].
  %%CITATION = PHRVA,D81,126009;%%

%\cite{Siopsis:2010uq}
\bibitem{Siopsis:2010uq}
  G.~Siopsis and J.~Therrien,
  %``Analytic calculation of properties of holographic superconductors,''
  JHEP {\bf 1005}, 013 (2010)
  [arXiv:1003.4275 [hep-th]].
  %%CITATION = JHEPA,1005,013;%%

%\cite{Li:2011xj}
\bibitem{Li:2011xj}
  H.~F.~Li, R.~G.~Cai and H.~Q.~Zhang,
  %``Analytical Studies on Holographic Superconductors in Gauss-Bonnet
  %Gravity,''
  arXiv:1103.2833 [hep-th].
  %%CITATION = ARXIV:1103.2833;%%

%\cite{Ge:2010aa}
\bibitem{Ge:2010aa}
  X.~-H.~Ge, B.~Wang, S.~-F.~Wu, G.~-H.~Yang,
  %``Analytical study on holographic superconductors in external magnetic field,''
  JHEP {\bf 1008}, 108 (2010).
  [arXiv:1002.4901 [hep-th]].

%\cite{Pan:2009xa}
\bibitem{Pan:2009xa}
  Q.~Pan, B.~Wang, E.~Papantonopoulos, J.~Oliveira, A.~B.~Pavan,
  %``Holographic Superconductors with various condensates in Einstein-Gauss-Bonnet gravity,''
  Phys.\ Rev.\  {\bf D81}, 106007 (2010).
  [arXiv:0912.2475 [hep-th]].

%\cite{Maeda:2008ir}
\bibitem{Maeda:2008ir}
  K.~Maeda and T.~Okamura,
  %``Characteristic length of an AdS/CFT superconductor,''
  Phys.\ Rev.\  D {\bf 78}, 106006 (2008)
  [arXiv:0809.3079 [hep-th]].
  %%CITATION = PHRVA,D78,106006;%%

%\cite{Kanno:2010pq}
\bibitem{Kanno:2010pq}
  S.~Kanno and J.~Soda,
  %``Stability of Holographic Superconductors,''
  Phys.\ Rev.\  D {\bf 82}, 086003 (2010)
  [arXiv:1007.5002 [hep-th]].
  %%CITATION = PHRVA,D82,086003;%%

%\cite{Brihaye:2010mr}
\bibitem{Brihaye:2010mr}
  Y.~Brihaye and B.~Hartmann,
  %``Holographic Superconductors in 3+1 dimensions away from the probe limit,''
  Phys.\ Rev.\  D {\bf 81}, 126008 (2010)
  [arXiv:1003.5130 [hep-th]].
  %%CITATION = PHRVA,D81,126008;%%

%\cite{Barclay:2010up}
\bibitem{Barclay:2010up}
  L.~Barclay, R.~Gregory, S.~Kanno and P.~Sutcliffe,
  %``Gauss-Bonnet Holographic Superconductors,''
  JHEP {\bf 1012}, 029 (2010)
  [arXiv:1009.1991 [hep-th]].
  %%CITATION = JHEPA,1012,029;%%

%\cite{Pan:2011ns}
\bibitem{Pan:2011ns}
  Q.~Pan, B.~Wang,
  %``General holographic superconductor models with backreactions,''
  [arXiv:1101.0222 [hep-th]].




\end{thebibliography}
\end{document}